\documentclass[letterpaper, 10 pt, conference]{ieeeconf}  % Comment this line out if you need a4paper

\IEEEoverridecommandlockouts                              % This command is only needed if 
\usepackage{bm}
\usepackage{enumerate}
\usepackage{commath}
\usepackage{graphicx}
\graphicspath{{Pictures/}}
\usepackage{amsmath}
\usepackage{breqn}
\usepackage{cite}
\usepackage[table]{xcolor}
\usepackage{booktabs}
\usepackage{empheq}
\usepackage{amsfonts}
\usepackage{comment}
\usepackage{amssymb}

\usepackage{amsthm}
\usepackage{hyperref}
\usepackage{xcolor}
\usepackage{mathrsfs}
\usepackage{accents}
\usepackage{thmtools}
\usepackage{thm-restate}
\usepackage{float}
\usepackage{xcolor}
\usepackage{algorithm}
\usepackage{algpseudocode}
\usepackage{amsfonts}
\usepackage[normalem]{ulem}

\theoremstyle{plain}
\newtheorem{corollary}{Corollary}
\newtheorem{lemma}{Lemma}
\newtheorem{definition}{Definition}
\newtheorem{theorem}{Theorem}
\newtheorem{assumption}{Assumption}

\newtheorem{problem}{Problem}
\newtheorem*{problem*}{Problem}
\newtheorem*{theorem*}{Theorem}
\newtheorem{assumption*}{Assumption}

\newtheorem{remark}{Remark}

\theoremstyle{definition}
\title{\LARGE \bf
Robust Visual Servoing under Human Supervision for Assembly Tasks
}

\author{Victor Nan Fernandez-Ayala$^{1}$, Jorge Silva$^{1}$, Meng Guo$^{2}$ and Dimos V. Dimarogonas$^{1}$% <-this % stops a space
\thanks{*This work was supported by the the ERC CoG LEAFHOUND, the EU CANOPIES project, the Knut and Alice Wallenberg Foundation (KAW) and the Digital Futures Smart Construction project.}% <-this % stops a space
\thanks{$^{1}$Victor Nan Fernandez-Ayala, Jorge Silva and Dimos V. Dimarogonas are with the Division of Decision and Control Systems, School of EECS, Royal Institute of Technology (KTH), 100 44 Stockholm, Sweden (Email: 
        {\tt\small vnfa, jafs2, dimos@kth.se}).}%
\thanks{$^{2}$Meng Guo is with the College of Engineering, Peking University, 100871 Beijing, China. (Email: 
        {\tt\small meng.guo@pku.edu.cn}).}%
}

\begin{document}

\maketitle
\thispagestyle{empty}
\pagestyle{empty}

%%%%%%%%%%%%%%%%%%%%%%%%%%%%%%%%%%%%%%%%%%%%%%%%%%%%%%%%%%%%%%%%%%%%%%%%%%%%%%%%

\newcommand{\myvec}[1]{\boldsymbol{#1}}
\newcommand{\mymatrix}[1]{\boldsymbol{#1}}
\newcommand{\myframe}[1]{\mathcal{#1}}
\newcommand{\dt}{\frac{d}{dt}}

\begin{abstract}
    We propose a framework enabling mobile manipulators to reliably complete pick-and-place tasks for assembling structures from construction blocks. The picking uses an eye-in-hand visual servoing controller for object tracking with Control Barrier Functions (CBFs) to ensure fiducial markers in the blocks remain visible. An additional robot with an eye-to-hand setup ensures precise placement, critical for structural stability. We integrate human-in-the-loop capabilities for flexibility and fault correction and analyze robustness to camera pose errors, proposing adapted barrier functions to handle them. Lastly, experiments validate the framework on 6-DoF mobile arms. 
\end{abstract}

\section{Introduction}
The construction industry faces low automation due to unique, strictly regulated, one-off projects. While software tools enable incremental improvements, robotics offer significant potential to increase safety, efficiency, and sustainability. Research on autonomous robotic assembly has recently expanded. On the issue of task planning, notable works include \cite{multi_robot_rearrangement_planning} for multi-robot task planning with parallelization and synchronization and \cite{physics_informed} with a physics-based approach for assembly. Object pose estimation has also seen a lot of recent developments, e.g., \cite{Wen_2023_CVPR} which tackles the problem of 6-DoF object tracking and 3D reconstruction from an RGBD video. We complement these advancements by focusing on precise low-level planning and control for pick-and-place tasks. 

% On the issue of task planning, \cite{TL_smart_manufacturing_systems} proposes an LTL-based framework for flexible and safe manufacturing, \cite{multi_robot_rearrangement_planning} focuses on multi-robot task planning with parallelization and synchronization, and \cite{physics_informed} proposes a physics-based approach for assembly planning. Object pose estimation has also seen a lot of recent developments, with \cite{10.1109/ICRA.2019.8793744}  focusing on category-agnostic object segmentation using synthetic depth images, \cite{DBLP:journals/corr/abs-2003-03518} proposing a method for depth-based pose estimation of occluded objects and  \cite{Wen_2023_CVPR} tackling the problem of 6-DoF object tracking and 3D reconstruction from an RGBD video.

In this area, visual servoing has been extensively used for object manipulation tasks, e.g., \cite{Allen} where an eye-in-hand architecture was used to reliably estimate 3D parameters with an adaptive control scheme to grasp dynamic objects. In \cite{SAHA} deformable linear objects,  e.g., ropes, were considered with a collaborative motion planning method. Lastly, in \cite{ZHANG20234381} the authors mix active sensing and manipulation, proposing to use Control Barrier Functions (CBFs) together with an image-based visual servoing scheme to avoid occlusions. 

In this work, we employ the CBF approach for visual servoing using an eye-in-hand architecture to ensure the object remains detectable during picking operations, allowing to rapidly detect and adjust to any movements that the object may undertake, which can happen if it is being handed to the robot by a human or another robot. This differs from \cite{ZHANG20234381} since the visual servoing here is position-based and an analysis of the robustness of the CBFs is performed, focusing on errors in the pose of the camera relative to the robot, and proposing corrections to deal with them and robustify the approach. We also employ an eye-to-hand scheme on an auxiliary robot to avoid occlusions generated by the handled object by sending visual feedback about the relative pose between the object and the structure it will be appended to. This collaborative setup for placing allows for greater precision and eliminates the assumption that the exact pose of the structure is known.

A software implementation of these basic actions is provided, allowing integration into broader systems with minimal effort and providing flexibility to the user. In addition, since construction is a safety critical field where adaptability is key, provisions are made to allow for safe human intervention to the robots. The software is validated with experiments using omnidirectional 6-DoF mobile manipulators. 

\section{Preliminaries}

\subsection{Notation}
We represent points/vectors by $\myvec{k} \in \mathbb{R}^n$, reference frames by $\mathcal{F}$ and matrices by $\myvec{A} \in \mathbb{R}^{m\times n}$. This representation is frame dependent, i.e, $\prescript{\myframe{F}}{}{}\myvec{k}$ stands for  point/vector $\myvec{k}$ in $\myframe{F}$. We also denote by $\prescript{\myframe{F}}{}{}\myvec{R}_{\myframe{G}} = \prescript{\myframe{G}}{}{}\myvec{R}_{\myframe{F}}^{-1} = \prescript{\myframe{G}}{}{}\myvec{R}_{\myframe{F}}^{T}$ the rotation from $\myframe{F}$ to $\myframe{G}$, whose columns are the basis vectors of $\myframe{G}$ represented in $\myframe{F}$, and by $\prescript{\myframe{F}}{}{}\myvec{t}_{\myframe{G}}$ the vector from the origin of $\myframe{F}$ to $\myframe{G}$, in frame $\myframe{F}$. A reference frame is fully defined by the position of its origin and orientation, and in Cartesian coordinates is associated with three perpendicular basis vectors, labeled $x$, $y$ and $z$. $\myvec{k}_z$ denotes the $z$-element of $k$, and similarly for $x$ and $y$. The derivative of vector $\myvec{k}$ with respect to time is denoted $\dot{\myvec{k}}$. We will also refer to $\myframe{C}_t$ as the camera frame at time $t$, $\myframe{E}_t$ as the end-effector frame at time $t$ and $\myframe{E}^*$ as the end-effector target frame.

\subsection{Pinhole camera model} \label{subsec:pinhole}
The intrinsic parameters of a camera can be represented by the upper triangular matrix $\myvec{K} \in \mathbb{R}^{3 \times 3}$. A point $\myvec{q} \in \mathbb{R}^3$ projected onto the image plane of the camera, $\myvec{r} \in \mathbb{R}^2$, has the relation $[
\prescript{\myframe{I}}{}{}\myvec{r}^T \; 1]^T \propto \myvec{K} \prescript{\myframe{C}}{}{}\myvec{q}$, where $\propto$ denotes proportionality, $\myframe{C}$ is the camera frame and $\myframe{I}$ is the two dimensional coordinate frame of the image. $\myframe{C}$ is defined such that its $z$ basis vector is pointing outwards from the camera and is perpendicular to the image plane, and its $x$ and $y$ basis vectors are pointing right and upwards from the image plane, respectively. Its origin is located in the pinhole of the camera. The $x$ and $y$ basis vectors of $\myframe{I}$ are pointing right and down, respectively, with its origin located in the top left corner of the image. If $\prescript{\myframe{C}}{}{}\myvec{q}$ is known, we can uniquely solve for $\prescript{\myframe{I}}{}{}\myvec{r}$, but if only $\prescript{\myframe{I}}{}{}\myvec{r}$ is known, we can only determine the line along which lies point $\myvec{q}$, which is described by $\myvec{K}^{-1} [ \prescript{\myframe{I}}{}{}\myvec{r} \; 1
]^T \propto \prescript{\myframe{C}}{}{}\myvec{q}$. 

\subsection{Visual servoing}
Visual servoing techniques \cite{visual_servoing_basic_approaches} seek to minimize the error $\myvec{e} = \myvec{s} - \myvec{s}^*$, where $\myvec{s} \in \mathbb{R}^k$ is a vector of $k$ visual features extracted from image measurements, and $\myvec{s}^*$ is their desired value. In this work, we consider $\myvec{s}^*$ to be constant and end-effector dynamics $\dot{\myvec{x}} = \myvec{u}$, where $\myvec{x}$ is the end-effector position and orientation and $\myvec{u}$ is the control input. To enforce an exponential decay of the error, $\dot{\myvec{e}} = -\sigma \myvec{e}$ with gain $\sigma \in  \mathbb {R}^+$, one can use $\myvec{u} = [
    \myvec{v} \; \myvec{w}
]^T = -\sigma \myvec{L_s}^\dagger \myvec{e}$, where $\myvec{L_s}^\dagger \in \mathbb{R}^{6 \times k}$ is the pseudo inverse of the interaction matrix such that $\dot{ \myvec{s}} = \myvec{L_s} [
    \myvec{v} \; \myvec{w}
]^T$, and $\myvec{v}$ and $\myvec{w}$ are the linear and angular velocity. $\myvec{L_s}$ depends on the choice of features $\myvec{s}$. We define $\boldsymbol{s} = [ \prescript{\myframe{E}^*}{}{}\boldsymbol{t}_{\myframe{E}_t} \; \myvec{\theta}\myvec{b} ]^T$, where $\myvec{\theta} \myvec{b}$ is the rotation vector representation of $\prescript{\myframe{E}^*}{}{}\myvec{R}_{\myframe{E}_t}$. Following \cite{visual_servoing_basic_approaches}, this $\myvec{s}$ leads to
\begin{equation*}
    \mymatrix{L_s} = \begin{bmatrix}
        \prescript{\myframe{E}^*}{}{}\mymatrix{R}_{\myframe{E}_t} & \mymatrix{0} \\
        \mymatrix{0} & \mymatrix{L_{\theta \myvec{b}}}
    \end{bmatrix},
\end{equation*}

\noindent where $
    \mymatrix{L_{\myvec{\theta} \myvec{b}}} := \mymatrix{I} - \frac{\myvec{\theta}}{2} [\myvec{b}]_\times + \left( 1 - \frac{\text{sinc}(\myvec{\theta})}{\text{sinc}^2(\myvec{\theta} / 2) } \right) [\myvec{b}]^2_\times
$ and $\text{sinc}(x)$ is the sinus cardinal function such that
$\text{sinc}(x)x = \text{sin}(x)$ and $\text{sinc}(0) = 1$. This leads to the control law
\begin{equation} \label{eq:visual_servoing_p_controller}
    \begin{cases}
  \myvec{v} = -\sigma \prescript{\myframe{E}^*}{}{}\mymatrix{R}_{\myframe{E}_t}^T \prescript{\myframe{E}^*}{}{}\boldsymbol{t}_{\myframe{E}_t} \\
  \myvec{w} = -\sigma \myvec{\theta} \myvec{b} 
  \end{cases}
\end{equation}
For further derivation details of \eqref{eq:visual_servoing_p_controller} we refer readers to \cite{visual_servoing_basic_approaches}. Note that \eqref{eq:visual_servoing_p_controller} requires a continuous estimate of $\prescript{\myframe{E}^*}{}{}\myvec{R}_{\myframe{E}_t}$ and $\prescript{\myframe{E}^*}{}{}\myvec{t}_{\myframe{E}_t}$. While complex algorithms for object pose estimation exist \cite{Wen_2023_CVPR}, in this work fiducial markers attached to the objects will be used. Specifically, we will use ArUco markers whose 6D pose estimation algorithm can be found in \cite{aruco_paper}.

\subsection{Control barrier functions} \label{subsec:cbf}
 Control barrier functions (CBFs) are based on the concept of a safety set $\myframe{S}$ which can be defined through a differentiable function of the system state, $h: \mathbb{R}^n \xrightarrow{} \mathbb{R}$ as follows
\begin{equation} \label{eq:set_s}
    \mathcal{S} := \{\myvec{x} \in \mathbb{R}^n \ | \ h(\boldsymbol{x}) \geq 0\},
\end{equation}

\noindent where $h$ can be thought of as a safety margin, with a value of zero on the boundary of the safe set. To formally define CBFs, we must first introduce the following notions:

\begin{comment}
\begin{definition}[Locally Lipshitz function] A function $f: \mathcal{D} \xrightarrow{} \mathcal{Y}$ is called locally Lipschitz if at each $\boldsymbol{x} \in \mathcal{D}$, there exists $\delta > 0$ such that 

\begin{equation*}
\text{sup}\{ \frac{\rho(f(a),f(b))}{d(a,b)} : a \neq b \; \wedge \{a, b\} \subseteq B_{\delta}(\myvec{x}) \} < \infty
\end{equation*}

\noindent, where $\rho$ and $d$ are distance metrics of spaces $\mathcal{D}$ and $\mathcal{Y}$, respectively, and $B_{\delta}(\myvec{x})$ is an open ball with center $\myvec{x}$ and radius $\delta$.

\end{definition}
\end{comment}

\begin{definition}[Extended class $\mathcal{K}$ function] \label{def:ext_K} A continuous increasing function $\alpha:[0,a) \rightarrow [0,\infty)$ with $\alpha(0)=0$.
\end{definition}

\begin{definition}[CBF] \label{def:cbf}
Let set $\mathcal{S}$ be defined by \eqref{eq:set_s}. $h(\myvec{x})$ is a CBF for the system $\dot{ \myvec{x}} = \boldsymbol{f}(\myvec{x}) + \boldsymbol{g}(\myvec{x})\myvec{u}$ if there exists a locally Lipschitz extended class $\mathcal{K}$ function $\alpha$ such that
\begin{align} \label{eq:cbf}
    \sup_{\myvec{u} \in \mathcal{U} \subseteq \mathbb{R}^m} & [L_{\myvec{f}}h(\myvec{x}) + L_{\myvec{g}}h(\myvec{x})\myvec{u} + \alpha(h(\myvec{x}))] \ge 0, \\ & \forall \myvec{x} \in \mathcal{D} \supseteq \mathcal{S} \subseteq \mathbb{R}^n \nonumber.
\end{align}
\end{definition}

Additionally, it has been shown \cite{Ames2017} that any locally Lipschitz control input $\myvec{u}$ that satisfies the CBF constraint \eqref{eq:cbf} renders the set $\mathcal{S}$ forward invariant and, if $\mathcal{S}$ is compact, it is also asymptotically stable. Since constraint \eqref{eq:cbf} is affine in $\myvec{u}$, a controller that satisfies it can be obtained by solving the quadratic problem (QP)
\begin{align} \label{eq:safety_set}
    \underset{\myvec{u} \in \mathbb{R}^m}{\text{min}} & \quad ||\myvec{u} - \myvec{u}_{\text{nom}}||_2^2 \nonumber \\
    \text{s.t.   } & \quad L_{\myvec{f}}h(\myvec{x}) + L_{\myvec{g}}h(\myvec{x})\myvec{u} + \alpha(h(\myvec{x})) \ge 0, 
\end{align}

\noindent where $\myvec{u}_{\text{nom}}$ is the input provided by an independent controller, e.g., \eqref{eq:visual_servoing_p_controller}. $\mathcal{S}$ can be represented as an intersection of $N$ other sets, $\mathcal{S} = \bigcap_{i=0}^{N-1} \mathcal{S}_i$, each defined by its own barrier function $h_i(\myvec{x})$, as in \eqref{eq:set_s}. In this case, one approach to ensure the forward invariance of $\mathcal{S}$ is ensuring that all barrier conditions are satisfied simultaneously as parallel constraints.

\section{Main results}
In this work, we enhance \eqref{eq:visual_servoing_p_controller} using CBFs for continuous marker detection, robustifying them against camera pose uncertainties. Additionally, we will adapt the algorithm for safe human control of the end-effector following \cite{victor_icra_hil}.

\subsection{Field of view}

\begin{figure}
    \centering
    % First column with two images stacked, labeled (a)
    \begin{minipage}{0.5\linewidth}
        \centering
        \includegraphics[width=\textwidth]{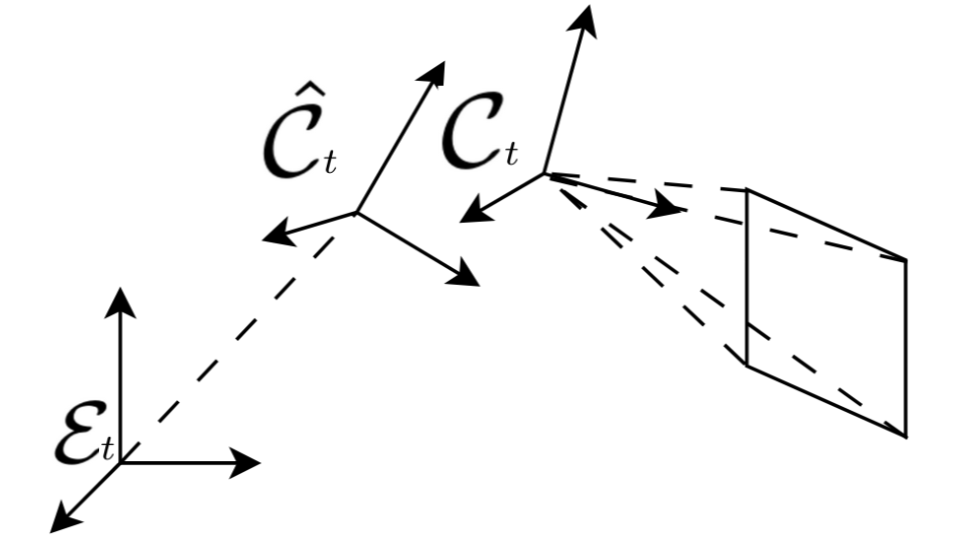}
    \end{minipage}%
    % Second column with one image, labeled (b)
    \begin{minipage}{0.5\linewidth}
        \centering
        \includegraphics[width=\textwidth]{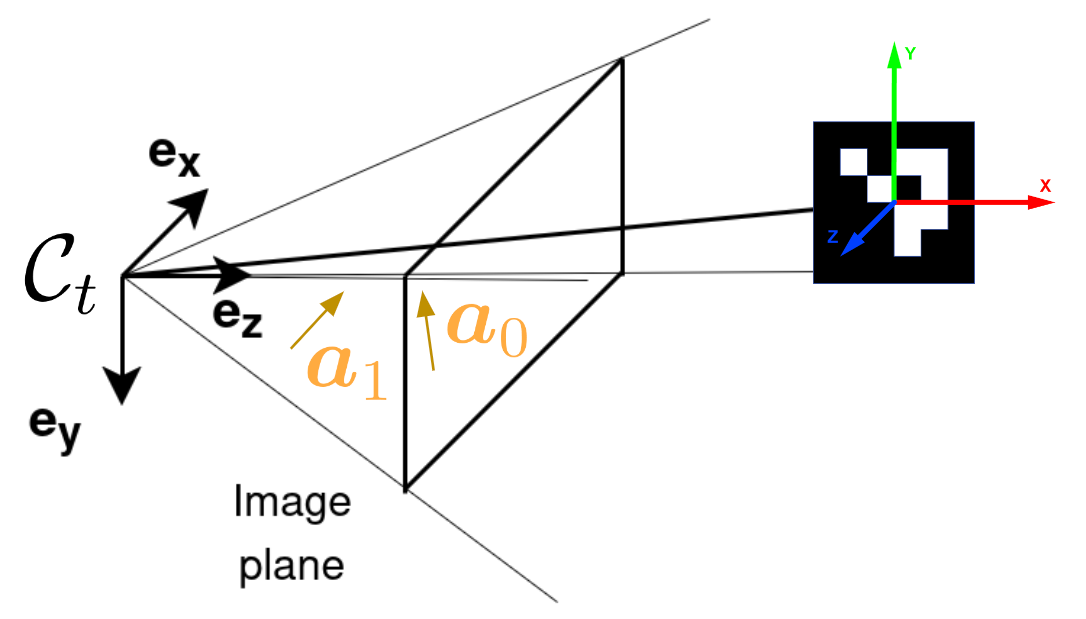}
    \end{minipage}
    \caption{Frames $\myframe{E}_t$, $\hat{\myframe{C}}_t$, $\myframe{C}_t$ and image plane (left). Dotted lines show known transformations and distances. Field of view and ArUco marker (right).}
    \label{fig:c_hat_and_c}
\end{figure}

The field of view of the camera, shown in Fig. \ref{fig:c_hat_and_c}, is defined as the set of 3D points that lie inside the volume enclosed by four \textit{visibility planes}. The visibility plane $i$ is defined as the set of points satisfying $\{\boldsymbol{y} \in \mathbb {R}^3 | \prescript{\myframe{C}_t}{}{}\boldsymbol{y} ^T \prescript{\myframe{C}_t}{}{}\boldsymbol{a}_i = 0\}$, where we set $||\boldsymbol{a}_i||_2 = 1$ with $\boldsymbol{a}_i$ pointing \textbf{into} the field of view. As such, we can define the field of view of the camera as
\begin{equation} \label{eq:volume_equation}
    \myframe{V}_{\myframe{C}_t} := \{\myvec{y} \in \mathbb{R}^3 \ | \ \forall i \in \{0, 1, 2, 3\},\prescript{\myframe{C}_t}{}{}\myvec{y}^T \prescript{\myframe{C}_t}{}{}\myvec{a}_i \geq 0 \}.
\end{equation} 

To determine the vectors $\myvec{a}_i$, we note that if the image has a length of $L$ and a width of $W$ (in pixels), its corners are given by $\prescript{I}{}{}\myvec{c}_0 := [
    0 \; 0
]^T$, $\prescript{I}{}{}\myvec{c}_1 := [
    0 \; L
]^T$, $\prescript{I}{}{}\myvec{c}_2 := [
    W \; L
]^T$, $\prescript{I}{}{}\myvec{c}_3 := [
    W \; 0
]^T$ (recall $\mathcal{I}$ basis vectors $x$ and $y$ are pointing right and down). Next, let the 3D line from the pinhole of the camera to the 3D location of image corner $i$ be the following collection of points $\{\Tilde{\boldsymbol{y}} \in \mathbb {R}^3 \ | \ \lambda \prescript{\myframe{C}_t}{}{}\boldsymbol{l}_i = \prescript{\myframe{C}_t}{}{}\Tilde{\boldsymbol{y}}, \lambda \in  \mathbb {R}^+ \}$, with $||\myvec{l}_i|| = 1$ such that $\prescript{\myframe{C}_t}{}{}l_{i,z} > 0$. To determine $\myvec{l}_i$, we note that $\myvec{d}_i := \myvec{K
}^{-1} [
    \prescript{\myframe{I}}{}{}\myvec{c}_i \; 1 
] \propto \prescript{\myframe{C}_t}{}{}\myvec{l}_i$. Given the aforementioned restrictions on $\myvec{l}_i$, we get $\prescript{\myframe{C}_t}{}{}\myvec{l}_i =  \myvec{d}_i / ||\myvec{d}_i||_2\cdot \text{sign}(d_{i, z}), \text{for } i = 0, 1, 2, 3$. Vectors $\myvec{a}_i$ are defined by $\myvec{a}_i = -\myvec{l}_i \times \myvec{l}_{(i+1)\text{mod}4}$ for $i = 0, 1, 2, 3$,  where the minus sign was introduced to ensure that each $\myvec{a}_i$ is pointing into the field of view. Note we are going \textit{around} the corners of the image, hence we use $\myvec{l}_{(i+1)\text{mod}4}$. We also make explicit the dependency of the field of view on the pose of the camera frame $\myframe{C}_t$ and the parameters $\myvec{K}, \myvec{C}$, i.e, $\myframe{V}_{\myframe{C}_t} = \myframe{V}_{\myframe{C}_t}\left( \myvec{K}, \myvec{C}\right)$, where $\myvec{C} := [
    \prescript{\myframe{I}}{}{}\myvec{c}_0 \; \prescript{\myframe{I}}{}{}\myvec{c}_1 \; \prescript{\myframe{I}}{}{}\myvec{c}_2 \;\prescript{\myframe{I}}{}{}\myvec{c}_3 \;
]$.

\subsection{CBFs for continued pose estimation}
To have uninterrupted detection of the marker, the four corners must always remain inside the field of view of the camera. This can be done by using the barrier functions
\begin{align} \label{eq:cube_barrier_function}
    h_{ij}(\prescript{\myframe{C}_t}{}{}\boldsymbol{x}_j) = \prescript{\myframe{C}_t}{}{}\boldsymbol{a}_i ^T \prescript{\myframe{C}_t}{}{}\boldsymbol{x}_j, \text{ for } i, j \in \{0, 1, 2, 3\},
\end{align}

\noindent where $\myvec{x}_j \in \mathbb{R}^3$ is the position of the $j$-th ArUco corner and $h_{ij}$ is the distance from corner $j$ to visibility plane $i$. If all CBFs are enforced simultaneously, then $\forall t > 0, i, j \in \{0, 1, 2, 3\}: h_{ij}(\prescript{\myframe{C}_t}{}{}\myvec{x}_j) > 0$, which mean all corners remain inside the field of view of the camera. While \textit{necessary}, this condition is not \textit{sufficient} for ensured continued detectability, since if the camera is \textit{behind} the marker and pointing in its direction, it is still possible that all CBFs are positive. To account for this possibility, we must add an extra constraint $h_z(\prescript{\myframe{A}}{}{}t_{\myframe{C}_t, z}) = \prescript{\myframe{A}}{}{}t_{\myframe{C}_t, z} - \zeta$, where $\myframe{A}$ refers to the ArUco. This enforces the camera stays in front of the marker by $\zeta \in \mathbb{R}_+$. 

\subsection{System dynamics of the CBFs}
To arrive at the constraints for the previous CBFs, we must know their dynamics in relation to the control input, i.e, the end-effector twist as represented in the end-effector frame, $\prescript{\myframe{E}_t}{}{}\myvec{\myvec{u}} = [
    \prescript{\myframe{E}_t}{}{}\myvec{v} \; \prescript{\myframe{E}_t}{}{}\myvec{w}
]$. For the barrier functions in \eqref{eq:cube_barrier_function}, we first use the relation between the velocity of a moving reference frame, i.e, the end-effector, and the velocity of a stationary point, i.e, ArUco corner $i$, as seen from the moving frame:
\begin{align*}
    \prescript{\myframe{E}_t}{}{} \dot{\boldsymbol{x}}_i = [\prescript{\myframe{E}_t}{}{}\boldsymbol{w}]_{\times}^T \prescript{\myframe{E}_t}{}{} \boldsymbol{x}_i - \prescript{\myframe{E}_t}{}{} \boldsymbol{v} = - [\prescript{\myframe{E}_t}{}{} \boldsymbol{x}_i]_{\times}^T \prescript{\myframe{E}_t}{}{} \boldsymbol{w} - \prescript{\myframe{E}_t}{}{} \boldsymbol{v}.
\end{align*}

To write the previous relation in terms of the CBFs state variables, we perform the coordinate transformation (translation and rotation) from end-effector to camera frame
\begin{align*} 
% \label{dynamics_cube_cam_frame}
    \prescript{\myframe{C}_t}{}{} \dot{\boldsymbol{x}}_i = - \prescript{\myframe{E}_t}{}{}\myvec{R}_{\myframe{C}_t}^T [\prescript{\myframe{E}_t}{}{}t_{\myframe{C}_t} + \prescript{\myframe{E}_t}{}{}\myvec{R}_{\myframe{C}_t} \prescript{\myframe{C}_t}{}{}\boldsymbol{x}_i]_{\times}^T \prescript{\myframe{E}_t}{}{}\boldsymbol{w} - \prescript{\myframe{E}_t}{}{}\myvec{R}_{\myframe{C}_t}^T \prescript{\myframe{E}_t}{}{}\boldsymbol{v}.
\end{align*}

For the extra constraint, note that $\prescript{\myframe{A}}{}{}\myvec{t}_{\myframe{C}_t} = \sum_k \prescript{\myframe{E}_t}{}{}t_{\myframe{C}_t, k} \cdot \prescript{\myframe{A}}{}{}\myvec{i}_k^{\myframe{E}_t}$ where $\myvec{i}_k^{\myframe{E}_t}$ are the basis vectors of $\myframe{E}_t$, i.e., $k=\{x, y, z\}$. The linear velocity of $\myframe{C}_t$ induced by $\myframe{E}_t$ rotational motion is
\begin{align*}
    & \left(  \prescript{\myframe{A}}{}{}\dot{\myvec{t}}_{\myframe{C}_t}\right)_{\text{rot}}   = \sum_k \prescript{\myframe{E}_t}{}{}t_{\myframe{C}_t, k} \left( \prescript{\myframe{A}}{}{}\myvec{w} \times \prescript{\myframe{A}}{}{}\myvec{i}_k^{\myframe{E}_t} \right) \\
    & = \sum_k \prescript{\myframe{E}_t}{}{}t_{\myframe{C}_t, k} \left( \prescript{\myframe{C}_t}{}{}\myvec{R}_{\myframe{A}}^T \prescript{\myframe{E}_t}{}{}\myvec{R}_{\myframe{C}_t}^T \prescript{\myframe{E}_t}{}{}\myvec{w} \times \prescript{\myframe{C}_t}{}{}\myvec{R}_{\myframe{A}}^T \prescript{\myframe{E}_t}{}{}\myvec{R}_{\myframe{C}_t}^T \prescript{\myframe{E}_t}{}{}\myvec{i}_k^{\myframe{E}_t} \right).
\end{align*}

The contribution of the linear velocity of the end-effector to the linear velocity of the camera is given by $( \prescript{\myframe{A}}{}{}\dot{\myvec{t}}_{\myframe{C}_t})_{\text{lin}} = \prescript{\myframe{C}_t}{}{}\myvec{R}_{\myframe{A}}^T \prescript{\myframe{E}_t}{}{}\myvec{R}_{\myframe{C}_t}^T \prescript{\myframe{E}_t}{}{}\myvec{v}$, with the total linear velocity of the camera being $ \prescript{\myframe{A}}{}{}\dot{\myvec{t}}_{\myframe{C}_t}= ( \prescript{\myframe{A}}{}{}\dot{\myvec{t}}_{\myframe{C}_t})_{\text{rot}} + ( \prescript{\myframe{A}}{}{}\dot{\myvec{t}}_{\myframe{C}_t})_{\text{lin}}.$ Since we consider end-effector dynamics $\dot{\myvec{x}} = \myvec{u}$, the constraint for each $h_{ij}$ in \eqref{eq:cube_barrier_function}, after transformation back to the end-effector frame, becomes 
\begin{equation} \label{eq:vis_barrierconditions}
\begin{aligned}
     \begin{bmatrix}
        \boldsymbol{V}_i(\myvec{K}, \myvec{C}, \myframe{C}_t) & \boldsymbol{W}_{ij} (\myvec{K}, \myvec{C}, \myframe{C}_t)
    \end{bmatrix} \prescript{\myframe{E}_t}{}{}\boldsymbol{\myvec{u}} \\
    + \boldsymbol{m}_{ij}(\myvec{K}, \myvec{C}, \myframe{C}_t) \geq 0,
\end{aligned}
\end{equation}

\noindent where $\boldsymbol{V}_i(\myvec{K},\myvec{C}, \myframe{C}_t) := - \prescript{\myframe{C}_t}{}{}\boldsymbol{a}_i^T \prescript{\myframe{E}_t}{}{}\myvec{R}_{\myframe{C}_t}^T$, $\boldsymbol{W}_{ij}(\myvec{K}, \myvec{C}, \myframe{C}_t) := - \prescript{\myframe{C}_t}{}{}\boldsymbol{a}_i^T\prescript{\myframe{E}_t}{}{}\myvec{R}_{\myframe{C}_t}^T [\prescript{\myframe{E}_t}{}{}\myvec{t}_{\myframe{C}_t} + \prescript{\myframe{E}_t}{}{}\myvec{R}_{\myframe{C}_t} \prescript{\myframe{C}_t}{}{}\boldsymbol{x}_j]_{\times}^T$ and $\boldsymbol{m}_{ij}(\myvec{K}, \myvec{C}, \myframe{C}_t) := \alpha (\prescript{\myframe{C}_t}{}{}\boldsymbol{a}_i ^T \prescript{\myframe{C}_t}{}{}\boldsymbol{x}_j)$. For the extra constraint, we use the total linear velocity of $\myframe{C}_t$ previously derived and the constraint becomes
\begin{equation} \label{eq:z_barrierconditions}
    \begin{bmatrix}
        \begin{bmatrix}
        0 & 0 & 1
    \end{bmatrix}\myvec{M} & \begin{bmatrix}
        0 & 0 & 1
    \end{bmatrix}\myvec{N}
    \end{bmatrix} \prescript{\myframe{E}_t}{}{}\myvec{\myvec{u}} + \myvec{n} \geq 0,
\end{equation}

\noindent where $\myvec{M} := \prescript{\myframe{\myframe{C}}_t}{}{}\myvec{R}_{\myframe{A}}^T \prescript{\myframe{E}_t}{}{}\myvec{R}_{\myframe{C}_t}^T$, $\myvec{n} := \alpha \left( \prescript{\myframe{A}}{}{}t_{\myframe{C}_t, z} - \zeta \right)$ and $\myvec{N} := - \left( \sum_j \prescript{\myframe{E}_t}{}{}t_{\myframe{C}_t, j} [\prescript{\myframe{C}_t}{}{}\myvec{R}_{\myframe{A}}^T \prescript{\myframe{E}_t}{}{}\myvec{R}_{\myframe{C}_t}^T \prescript{\myframe{E}_t}{}{}\myvec{i}_j^{\myframe{E}_t}]_{\times} \right)  \prescript{\myframe{C}_t}{}{}\myvec{R}_{\myframe{A}}^T \prescript{\myframe{E}_t}{}{}\myvec{R}_{\myframe{C}_t}^T$. 

\begin{comment}
\begin{align*}
    \myvec{M} & := \prescript{\myframe{\myframe{C}}_t}{}{}\myvec{R}_{\myframe{A}}^T \prescript{\myframe{E}_t}{}{}\myvec{R}_{\myframe{C}_t}^T \\
    \myvec{N} & := - \left( \sum_j \prescript{\myframe{E}}{}{}t_{\myframe{C}_t, j} [\prescript{\myframe{\myframe{C}}}{}{}\myvec{R}_{\myframe{A}}^T \prescript{\myframe{E}_t}{}{}\myvec{R}_{\myframe{C}_t}^T \prescript{\myframe{E}}{}{}\myvec{i}_j^{\myframe{E}_t}]_{\times} \right)  \prescript{\myframe{\myframe{C}}}{}{}\myvec{R}_{\myframe{A}}^T \prescript{\myframe{E}_t}{}{}\myvec{R}_{\myframe{C}_t}^T  \\
    \myvec{n} & := \alpha \left( \prescript{\myframe{A}}{}{}t_{\myframe{C}_t, z} - \text{tol}_z \right)
\end{align*}
\end{comment}

\subsection{Robustifying the system} \label{subsec:robust_cbfs}

If \eqref{eq:vis_barrierconditions} and \eqref{eq:z_barrierconditions} are enforced, the system is only guaranteed to stay in the safe set if the precise pose of the camera with respect to the end-effector, $\prescript{\myframe{E}_t}{}{}\myvec{R}_{\myframe{C}_t}$, is known.  If there are errors, we must use an \textit{estimated} camera frame, $\hat{\myframe{C}}_t$, of which we know the pose, and an \textit{actual} camera frame, $\myframe{C}_t$, whose pose we do not know. In this case, we also have two different fields of view, one associated to $\myframe{C}_t$, $\myframe{V}_{\myframe{C}_t}(\myvec{K}, \myvec{C})$, and another one to $\hat{\myframe{C}}_t$, $\myframe{V}_{\hat{\myframe{C}}_t}(\hat{\myvec{K}}, \hat{\myvec{C}})$, where we associated $\hat{\myvec{K}}$ and $\hat{\myvec{C}}$ to frame $\hat{\myframe{C}}_t$ for the sake of generality. These errors mean we cannot enforce \eqref{eq:vis_barrierconditions} as that would require knowledge of the pose of $C_t$. If instead we enforce barrier conditions
\begin{equation} \label{eq:barrier_conditions_wdiff_params}
\begin{aligned}
%\label{eq:vis_barrierconditions}
     \begin{bmatrix}
        \boldsymbol{V}_i(\Tilde{\myvec{K}}, \Tilde{\myvec{C}}, \Tilde{\myframe{C}}_t) & \boldsymbol{W}_{ij} (\Tilde{\myvec{K}}, \Tilde{\myvec{C}}, \Tilde{\myframe{C}}_t)
    \end{bmatrix} \prescript{\myframe{E}_t}{}{}\boldsymbol{\myvec{u}} \\  + \boldsymbol{m}_{ij}(\Tilde{\myvec{K}}, \Tilde{\myvec{C}}, \Tilde{\myframe{C}}_t) \geq 0,
\end{aligned}
\end{equation}

\noindent then $\forall t, \ \forall j \in \{0, 1, 2, 3\}, \ \myvec{x}_j \in \myframe{V}_{\Tilde{\myframe{C}}_t}(\Tilde{\myvec{K}}, \Tilde{\myvec{C}})$. However, we want that $\forall t, \ \forall j \in \{0, 1, 2, 3\}, \ \myvec{x}_j \in \myframe{V}_{\myframe{C}_t}(\myvec{K}, \myvec{C})$. The former implies the latter if $\myframe{V}_{\Tilde{\myframe{C}}_t}(\Tilde{\myvec{K}}, \Tilde{\myvec{C}}) \subseteq \myframe{V}_{\myframe{C}_t}(\myvec{K}, \myvec{C})$. This condition is equivalent to having the visibility lines that enclose $\myframe{V}_{\Tilde{\myframe{C}}_t}(\Tilde{\myvec{K}}, \Tilde{\myvec{C}})$ fully contained in $\myframe{V}_{\myframe{C}_t}(\myvec{K}, \myvec{C})$: 
\begin{align} \label{eq:condtions_robustness}
    (\prescript{\myframe{C}_t}{}{}\myvec{t}_{\Tilde{\myframe{C}}_t} + \lambda \prescript{\myframe{C}_t}{}{}\myvec{R}_{\Tilde{\myframe{C}}_t} \prescript{\hat{\myframe{C}}_t}{}{}\Tilde{\myvec{l}}_j)^T \prescript{\myframe{C}_t}{}{}\myvec{a}_i \geq 0,  i, j \in \{0, 1, 2, 3\}, \lambda \in \mathbb{R}^{+}. 
\end{align}

\begin{remark}
\eqref{eq:z_barrierconditions} is already robust under errors in the camera pose by definition since it only involves the direction of the field of view (forward or backwards from the camera).
\end{remark}

\begin{assumption} \label{ass:general_problem}
    The camera frame pose error is bounded by an arbitrary quantity $\delta$ in translation and $\epsilon$ in rotation, i.e., $||\prescript{\myframe{C}_t}{}{}\myvec{t}_{\hat{\myframe{C}}_t}||_{2} \leq \delta$ and rotation $\prescript{\myframe{C}_t}{}{}\myvec{R}_{\hat{\myframe{C}}_t}$ is such that its rotation vector representation, $\myvec{\theta}\myvec{u}$, satisfies $||\myvec{\theta}\myvec{u}||_2 \leq \epsilon$.
\end{assumption}

We are now ready to state the main problem treated in this paper, regarding robustifying for errors in the camera pose.

\begin{problem} \label{prob:general_problem}
\begin{comment}
Suppose now that one can provide an upper bound to the dissimilarities between the poses of $\hat{\myframe{C}}_t$ and $\myframe{C}_t$. Concretely, assume \ref{ass:general_problem}. 
\end{comment}
Let $\myframe{P}(\delta, \epsilon)$ be the set of frames $\myframe{C}_t$ that satisfy Assumption \ref{ass:general_problem}. Suppose that we apply a rigid transformation to frame $\hat{\myframe{C}}_t$ and denote the result $\Tilde{\myframe{C}}_t$. The problem of robustifying barrier functions \eqref{eq:vis_barrierconditions} can be stated as the problem of finding the frame $\Tilde{\myframe{C}}_t$ and parameters $\Tilde{\myvec{K}}, \Tilde{\myvec{C}}$ such that
\begin{equation} \label{eq:problem_formulation}
\myframe{V}_{\Tilde{\myframe{C}}_t}(\Tilde{\myvec{K}}, \Tilde{\myvec{C}}) \subseteq \myframe{V}_{\myframe{F}_t}(\myvec{K}, \myvec{C}), \; \forall \myframe{F}_t \in \myframe{P}(\delta, \epsilon).
\end{equation}
\end{problem}

We split this problem into two simpler ones (Problem \ref{prob:zero_rotation_error} and Problem \ref{prob:zero_translation_error}) and use the findings to solve Problem \ref{prob:general_problem}. 

% \subsubsection{Zero rotation error}

\begin{comment}
\begin{assumption} \label{ass:zero_rotation}
    $||\prescript{\myframe{C}_t}{}{}\myvec{t}_{\hat{\myframe{C}}_t}||_{2} \leq \delta$ and $\prescript{\myframe{C}_t}{}{}\myvec{R}_{\hat{\myframe{C}}_t} = \mathbb{I}$
\end{assumption}
\end{comment}

\begin{problem} \label{prob:zero_rotation_error}
    Solve Problem \ref{prob:general_problem} for $\epsilon=0$.  
\end{problem}

\begin{comment}
In this sub-sub-section we assume that $||\prescript{\myframe{C}_t}{}{}\myvec{t}_{\hat{\myframe{C}}_t}||_{2} \leq \delta$ and that $\prescript{\myframe{C}_t}{}{}\myvec{R}_{\hat{\myframe{C}}_t} = \mathbb{I}$. We are now looking for a frame $\Tilde{\myframe{C}}_t$ and parameters $\Tilde{\myvec{K}}, \Tilde{\myvec{C}}$ such that $\myframe{V}_{\Tilde{\myframe{C}}_t}(\Tilde{\myvec{K}}, \Tilde{\myvec{C}}) \in \myframe{V}_{\myframe{F}_t}(\myvec{K}, \myvec{C})$ for all $\myframe{F}_t \in \myframe{P}(\delta, 0)$. 
\end{comment}

Problem \ref{prob:zero_rotation_error} deals with the case where $\myframe{C}_t$ has the same orientation as $\hat{\myframe{C}}_t$ and only errors regarding the position of $\myframe{C}_t$ exist. We start by introducing a Lemma when \eqref{eq:condtions_robustness} is modified to include the condition $\epsilon=0$. We will then use it to prove Theorem \ref{theo:rotation_zero} and apply it to solve Problem \ref{prob:zero_rotation_error}. 

\begin{lemma} \label{lemma:zero_rotation_erorr}
    If $\prescript{\myframe{F}_t}{}{}\myvec{R}_{\hat{\myframe{C}}_t} = I$, where $I$ is the identity matrix, and 
    \begin{align} \label{eq:condition_lemma_zero_rotation_error}
        \prescript{\myframe{F}_t}{}{}\myvec{t}_{\Tilde{\myframe{C}}_t}^T \prescript{\myframe{F}_t}{}{}\myvec{a}_i  \geq  0, \; \forall i \in \{0, 1, 2, 3\}, \; \lambda \in \mathbb{R}^{+},
    \end{align}
    then  $\myframe{V}_{\Tilde{\myframe{C}}_t}(\myvec{K}, \myvec{C}) \subseteq \myframe{V}_{\myframe{F}_t}(\myvec{K}, \myvec{C})$. 
\end{lemma}

\begin{proof}
Recall that $\myframe{V}_{\Tilde{\myframe{C}}_t}(\Tilde{\myvec{K}}, \Tilde{\myvec{C}}) \subseteq \myframe{V}_{\myframe{F}_t}(\myvec{K}, \myvec{C})$ if condition \eqref{eq:condtions_robustness} holds (where $\myframe{C}_t$ should be substituted by $\myframe{F}_t$ and $\hat{\myframe{C}}_t$ should be substituted by $\Tilde{\myframe{C}}_t$). If $\prescript{\myframe{C}_t}{}{}\myvec{R}_{\Tilde{\myframe{C}}_t} = I$, then $\Tilde{\myvec{K}} = \myvec{K}$, $\Tilde{\myvec{C}} = \myvec{C}$ and condition \eqref{eq:condtions_robustness} becomes 
\begin{align*}
    \prescript{\myframe{F}_t}{}{}\myvec{t}_{\Tilde{\myframe{C}}_t}^T \prescript{\myframe{F}_t}{}{}\myvec{a}_i \geq  - \lambda \prescript{\myframe{F}_t}{}{} \myvec{l}_j^T \prescript{\myframe{F}_t}{}{}\myvec{a}_i, \; \forall i,j \in \{0, 1, 2, 3\}, \; \lambda \in \mathbb{R}^{+}.
\end{align*}

\begin{comment}
\begin{align*}
    \prescript{\myframe{F}_t}{}{}\myvec{t}_{\Tilde{\myframe{C}}_t}^T \prescript{\myframe{F}_t}{}{}\myvec{a}_i & \geq  - \lambda \prescript{\myframe{F}_t}{}{} \Tilde{\myvec{l}}_j^T \prescript{\myframe{F}_t}{}{}\myvec{a}_i, \; \forall i,j \in \{0, 1, 2, 3\}, \lambda \in \mathbb{R}^{+}.
\end{align*}

If furthermore $\Tilde{\myvec{K}} = \myvec{K}$ and $\Tilde{\myvec{C}} = \myvec{C}$, vector $\Tilde{\myvec{l}}_j$ becomes the same as $\myvec{l}_j$, and the condition is further reduced to
\end{comment}

For a fixed $i$, the right hand side achieves a maximum when $j$ is such that $\prescript{\myframe{F}_t}{}{} \myvec{l}_j^T \prescript{\myframe{F}_t}{}{}\myvec{a}_i = 0$. This follows from the observation that $\prescript{\myframe{F}_t}{}{} \myvec{l}_j^T \prescript{\myframe{C}_t}{}{}\myvec{a}_i \geq 0$, and equals zero when $j\in\{i, (i+1)_{\text{mod}4}\}$. As such we obtain \eqref{eq:condition_lemma_zero_rotation_error}.
\end{proof}

\begin{comment}
Lemma \ref{lemma:zero_rotation_erorr} means that if $\prescript{\myframe{F}_t}{}{}\myvec{R}_{\hat{\myframe{C}}_t} = \mathbb{I}$, then $\myframe{V}_{\Tilde{\myframe{C}}_t}(\myvec{K}, \myvec{C}) \in \myframe{V}_{\myframe{F}_t}(\myvec{K}, \myvec{C})$ if and only if the origin of $\Tilde{\myframe{C}}_t$ belongs to $\myframe{V}_{\myframe{F}_t}(\myvec{K}, \myvec{C})$.
\end{comment}

\begin{theorem}
\label{theo:rotation_zero}
If $\prescript{\hat{\myframe{C}}_t}{}{}\myvec{t}_{\Tilde{\myframe{C}}_t} = [
    0 \; 0 \; \Tilde{z}
]$ and $\Tilde{z}$ satisfies 
\begin{equation} \label{eq:theorem_rotation_zero}
    \Tilde{z} \geq \frac{\delta}{\underset{i \in \{0, 1, 2, 3\}}{\normalfont{\text{min}}} \prescript{\myframe{C}_t}{}{}a_{i, z} },
\end{equation}
then $\myframe{V}_{\Tilde{\myframe{C}}_t}(\myvec{K}, \myvec{C}) \subseteq \myframe{V}_{\myframe{F}_t}(\myvec{K}, \myvec{C}), \; \forall \myframe{F}_t \in \myframe{P}(\delta, 0)$. 
\end{theorem}

\begin{comment}
In order to solve problem \ref{prob:zero_rotation_error}, we must find $\Tilde{\myframe{C}}_t$ such that \ref{eq:condition_lemma_zero_rotation_error} holds for all $\myframe{F}_t$ such that $||\prescript{\myframe{F}_t}{}{}\myvec{t}_{\hat{\myframe{C}}_t}||_{2} \leq \delta$.
\end{comment}

\begin{proof}
\begin{comment}
If assumption holds, then, by Lemma \ref{lemma:zero_rotation_erorr}, $\myframe{V}_{\Tilde{\myframe{C}}_t}(\myvec{K}, \myvec{C}) \in \myframe{V}_{\myframe{F}_t}(\myvec{K}, \myvec{C})$ if and only if \ref{eq:condition_lemma_zero_rotation_error}. It furthermore follows that $\myframe{V}_{\Tilde{\myframe{C}}_t}(\myvec{K}, \myvec{C}) \in \myframe{V}_{\myframe{F}_t}(\myvec{K}, \myvec{C}) \; \forall \myframe{F}_t \in \myframe{P}(\delta, 0)$ if and only if \ref{eq:condition_lemma_zero_rotation_error} $\forall \myframe{F}_t \in \myframe{P}(\delta, 0)$.
\end{comment}

By Lemma \ref{lemma:zero_rotation_erorr}, $\myframe{V}_{\Tilde{\myframe{C}}_t}(\myvec{K}, \myvec{C}) \subseteq \myframe{V}_{\myframe{F}_t}(\myvec{K}, \myvec{C}), \; \forall \myframe{F}_t \in \myframe{P}(\delta, 0)$ if \eqref{eq:condition_lemma_zero_rotation_error} holds $\forall \myframe{F}_t \in \myframe{P}(\delta, 0)$. If \eqref{eq:condition_lemma_zero_rotation_error} holds $ \forall \myframe{F}_t \in \myframe{P}(\delta, 0)$ when $||\prescript{\myframe{F}_t}{}{}\myvec{t}_{\hat{\myframe{C}}_t}||_{2} = \delta$, then it also holds generally $\forall \myframe{F}_t \in \myframe{P}(\delta, 0)$. We can thus restrict the analysis to the boundary of the sphere. Each point $\myvec{r}$ is such that $||\prescript{\hat{\myframe{C}}_t}{}{}\myvec{r}||_2 = \delta$ has an associated field of view, which can be written as
\begin{align*}
    \myframe{V}(\myvec{r}) = \{  \myvec{y} \in \mathbb{R}^3 \ | \ \forall i \in \{0, 1, 2, 3\}, (\prescript{\hat{\myframe{C}}_t}{}{}\myvec{y} - \prescript{\hat{\myframe{C}}_t}{}{}\myvec{r})^T\prescript{\myframe{C}_t}{}{}\myvec{a}_i \geq 0\}.
\end{align*} 

\begin{comment}
Taking into account the symmetry of the problem, we can look for rigid transformations of $\hat{\myframe{C}}_t$ that amount to translations along its own $z$ axis.
\end{comment}

Let $\myvec{q} \in \mathbb{R}^3$ be $\prescript{\hat{\myframe{C}}_t}{}{}\myvec{q} = [
    0 \; 0 \; \Tilde{z}
]$. Point $\myvec{q} \in \myframe{V}(\myvec{r})\;  \forall \myvec{r}: ||\prescript{\hat{\myframe{C}}_t}{}{}\myvec{r}||_2 = \delta$ if $ \Tilde{z}  \geq \prescript{\hat{\myframe{C}}_t}{}{}\myvec{r}^T \prescript{\myframe{C}_t}{}{}\myvec{a}_i/\prescript{\myframe{C}_t}{}{}a_{i, z}, \; \forall i \in \{0, 1, 2, 3\}, \; \myvec{r}: ||\prescript{\hat{\myframe{C}}_t}{}{}\myvec{r}||_2 = \delta
$ which is equivalent to 
\[
\Tilde{z} \geq \underset{i \in \{0, 1, 2, 3\}, ||\prescript{\hat{\myframe{C}}_t}{}{}\myvec{r}||_2 = \delta}{\text{max}}    \prescript{\hat{\myframe{C}}_t}{}{}\myvec{r}^T \prescript{\myframe{C}_t}{}{}\myvec{a}_i/\prescript{\myframe{C}_t}{}{}a_{i, z}.
\]

Given any $\myvec{a}_i$, one can find a point $\myvec{r}$ in the sphere that matches its direction with $\prescript{\hat{\myframe{C}}_t}{}{}\myvec{r}^T \prescript{\myframe{C}_t}{}{}\myvec{a}_i = \delta$. Thus, we get \eqref{eq:theorem_rotation_zero}.
\end{proof}

By Theorem \ref{theo:rotation_zero}, if we choose the origin of $\Tilde{C}_t$, $\myvec{q}$, to be
\begin{equation} \label{eq:problem2_final_sol}
    \prescript{\hat{\myframe{C}}_t}{}{}\myvec{q} = \begin{bmatrix}
    0 & 0 & \frac{\delta}{\underset{i \in \{0, 1, 2, 3\}}{\text{min}} \prescript{\myframe{C}_t}{}{}a_{i, z} }
\end{bmatrix},
\end{equation}

\noindent and its orientation $\prescript{\hat{\myframe{C}}_t}{}{}\myvec{R}_{\Tilde{\myframe{C}}_t} = I$,  leaving the remaining parameters equal, $\Tilde{\myvec{K}} = \myvec{K}$, $\Tilde{\myvec{C}} = \myvec{C}$, then condition \eqref{eq:problem_formulation} is satisfied and Problem \ref{prob:zero_rotation_error} is solved. We move next to Problem \ref{prob:zero_translation_error}. 

% \subsubsection{Zero translation error}

\begin{comment}
\begin{assumption} \label{ass:zero_translation}
    $||\prescript{\myframe{C}_t}{}{}\myvec{t}_{\hat{\myframe{C}}_t}||_{2} = 0$ and $\prescript{\myframe{C}_t}{}{}\myvec{R}_{\hat{\myframe{C}}_t}$ is such that its rotation vector representation, $\theta\myvec{u}$, satisfies $||\theta\myvec{u}||_2 \leq \epsilon$.
\end{assumption}
\end{comment}

\begin{problem} \label{prob:zero_translation_error}
    Solve Problem \ref{prob:general_problem} for $\delta=0$. 
\end{problem}

\begin{comment}
In this sub-sub-section, we consider the case where $\prescript{\myframe{C}_t}{}{}\myvec{t}_{\hat{\myframe{C}}_t} = \myvec{0}$ and $\prescript{\myframe{C}_t}{}{}\myvec{R}_{\hat{\myframe{C}}_t}$ is such that its rotation vector representation is $\theta \myvec{u}$, where $||\theta\myvec{u}||_2  \leq \epsilon$. As before, we seek to find a frame $\Tilde{\myframe{C}}_t$ and parameters $\Tilde{\myvec{K}}$ and $\Tilde{\myvec{C}}$ such that $\myframe{V}_{\Tilde{\myframe{C}}_t}(\Tilde{\myvec{K}}, \Tilde{\myvec{C}}) \in \myframe{V}_{\myframe{F}_t}(\myvec{K}, \myvec{C}), \; \forall  \myframe{F}_t \in \myframe{P}(0, \epsilon)$. 
\end{comment}

Problem \ref{prob:zero_translation_error} deals with the case where the position of $\myframe{C}_t$ coincides with $\hat{\myframe{C}}_t$ and only errors with its orientation exist. We start by refining Theorem \ref{theo:zero_translation} for the case of Problem \ref{prob:zero_translation_error}. 

\begin{theorem} \label{theo:zero_translation}
    If $\Tilde{\myframe{C}}_t$ coincides with $\hat{\myframe{C}}_t$, $\Tilde{\myvec{K}} = \myvec{K}$ and $\Tilde{\myvec{C}}$ is chosen according to Algorithm \ref{alg:zero_translation}, then $\myframe{V}_{\Tilde{\myframe{C}}_t}(\myvec{K}, \Tilde{\myvec{C}}) \subseteq \myframe{F}_t, \; \forall \myframe{F}_t \in \myframe{P}(0, \epsilon)$.
\end{theorem}

\begin{algorithm}
\caption{Compute corner matrix}\label{alg:zero_translation}
\begin{algorithmic}[1]
\For{$j \in \{0, 1, 2, 3\}$}
\State $\prescript{\hat{\myframe{C}}_t}{}{}\myvec{o}_j \gets \begin{bmatrix}
    \prescript{\hat{\myframe{C}}_t}{}{}\hat{l}_{j, x} & - \prescript{\hat{\myframe{C}}_t}{}{}\hat{l}_{j, y} & 0
\end{bmatrix}^T$
\State $\prescript{\hat{\myframe{C}}_t}{}{}\myvec{o}_j \gets \epsilon \frac{\myvec{o}_j}{||\myvec{o}_j||_2}$
\State $\prescript{\hat{\myframe{I}}}{}{}\myvec{t}_j \gets \textsc{project}\left(\myvec{R}(\prescript{\hat{\myframe{C}}_t}{}{}\myvec{o}_j) \prescript{\hat{\myframe{C}}_t}{}{}\hat{\myvec{l}}_j\right)$
\State $r_j \gets ||\prescript{\hat{\myframe{I}}}{}{}\myvec{t}_j - \prescript{\hat{\myframe{I}}}{}{}\myvec{c}_j ||_2$
\EndFor
\State $\prescript{\hat{\myframe{I}}}{}{}\Tilde{\myvec{c}}_0 \gets \begin{bmatrix}
    \frac{\sqrt{2}}{2}r_0 & \frac{\sqrt{2}}{2}r_0
\end{bmatrix}^T$, $\hat{\prescript{\myframe{I}}{}{}}\Tilde{\myvec{c}}_1 \gets \begin{bmatrix}
    \frac{\sqrt{2}}{2}r_1 & L - \frac{\sqrt{2}}{2}r_1
\end{bmatrix}^T$, $\prescript{\hat{\myframe{I}}}{}{}\Tilde{\myvec{c}}_2 \gets \begin{bmatrix}
    W - \frac{\sqrt{2}}{2}r_2 & L - \frac{\sqrt{2}}{2}r_2
\end{bmatrix}^T$, $\prescript{\hat{\myframe{I}}}{}{}\Tilde{\myvec{c}}_3 \gets \begin{bmatrix}
    W - \frac{\sqrt{2}}{2}r_3 & \frac{\sqrt{2}}{2}r_3
\end{bmatrix}^T$
\State $\Tilde{\myvec{C}} \gets \begin{bmatrix}
    \prescript{\hat{\myframe{I}}}{}{}\Tilde{\myvec{c}}_0 & \prescript{\hat{\myframe{I}}}{}{}\Tilde{\myvec{c}}_1 & \prescript{\hat{\myframe{I}}}{}{}\Tilde{\myvec{c}}_2 & \prescript{\hat{\myframe{I}}}{}{}\Tilde{\myvec{c}}_3
\end{bmatrix}$
\begin{comment}
\State $\myvec{c}_0 \gets \begin{bmatrix}
    0 & 0
\end{bmatrix}^T$
\State $\myvec{c}_1 \gets \begin{bmatrix}
    0 & L
\end{bmatrix}^T$
\State $\myvec{c}_2 \gets \begin{bmatrix}
    W & L
\end{bmatrix}^T$
\State $\myvec{c}_3 \gets \begin{bmatrix}
    W & 0
\end{bmatrix}^T$
\For{$i \in \{0, 1, 2, 3\}$}
\State $\myvec{d}_i \gets \frac{\myvec{K}^{-1} \myvec{c}_i}{||\myvec{K}^{-1} \myvec{c}_i||_2}$
\State $\prescript{\myframe{C}_t}{}{}\myvec{l}_i \gets \myvec{d}_i \cdot \frac{d_{i, z}}{||d_{i, z}||}$
\State $\prescript{\myframe{C}_t}{}{}\myvec{a}_i \gets - \prescript{\myframe{C}_t}{}{}\myvec{l}_i \times \prescript{\myframe{C}_t}{}{}\myvec{l}_{(i+1)\text{mod}4}$
\EndFor
\State \textbf{return} $(\prescript{\myframe{C}_t}{}{}\myvec{a}_0, \prescript{\myframe{C}_t}{}{}\myvec{a}_1, \prescript{\myframe{C}_t}{}{}\myvec{a}_2, \prescript{\myframe{C}_t}{}{}\myvec{a}_3)$
\end{comment}
\end{algorithmic}
\end{algorithm} 

\begin{proof}
Consider the set of points obtained by applying to visibility vector $\hat{\myvec{l}}_i$ rotations whose vector satisfies $||\myvec{\theta} \myvec{u}||_2 \leq \epsilon$. This set is a sphere segment centered on the origin of $\hat{\myframe{C}}_t$ whose projection onto the image plane of $\hat{\myframe{C}}_t$ is a circle centered at the corresponding image corner. To find the radius of the projected circles, one can follow steps $1-6$ of Algorithm \ref{alg:zero_translation}, where $\myvec{o}_j$ is a vector of norm $\epsilon$ that is perpendicular to $\hat{\myvec{l}}_j$ and stands for the rotation vector associated to rotation matrix $\myvec{R}(\prescript{\hat{\myframe{C}}_t}{}{}\myvec{o}_j)$. Applying $\myvec{R}(\prescript{\hat{\myframe{C}}_t}{}{}\myvec{o}_j)$ to $\hat{\myvec{l}}_j$ results in a point whose projection onto the image plane (given by the function \textsc{Project}) belongs on the aforementioned circle, with its radius given by step $5$. $\Tilde{\myvec{C}}$ is then computed by steps $7-8$, which ensure the new corners and line segments that unite them in the image plane do not intersect the interior of any of the circles.
\end{proof}

\begin{theorem}\label{thm:problem1_sol}
If $\Tilde{\myframe{C}}_t$ is chosen as \eqref{eq:problem2_final_sol} with $\prescript{\hat{\myframe{C}}_t}{}{}\myvec{R}_{\Tilde{\myframe{C}}_t} = I$ and $\Tilde{\myvec{C}}$ is computed using Algorithm \ref{alg:zero_translation}, then \eqref{eq:problem_formulation} holds.
\end{theorem}

\begin{proof}
By Lemma \ref{lemma:zero_rotation_erorr} and Theorem \ref{theo:rotation_zero} we know how to robustify translation error $\delta$ such that $\myframe{V}_{\Tilde{\myframe{C}}_t}(\myvec{K}, \myvec{C}) \subseteq \myframe{V}_{\myframe{F}_t}(\myvec{K}, \myvec{C}), \ \forall \myframe{F}_t \in \myframe{P}(\delta, 0)$  and by Theorem \ref{theo:zero_translation} we also know how to robustify rotation error $\epsilon$ such that  $\myframe{V}_{\Tilde{\myframe{C}}_t}(\myvec{K}, \Tilde{\myvec{C}}) \subseteq \myframe{F}_t, \ \forall \myframe{F}_t \in \myframe{P}(0, \epsilon)$, which when combined become \eqref{eq:problem_formulation}.
\end{proof}

Theorem \ref{thm:problem1_sol} can be used to solved Problem \ref{prob:general_problem} by finding $\Tilde{\myframe{C}}_t$ and $\Tilde{\myvec{K}}, \Tilde{\myvec{C}}$ satisfying \eqref{eq:problem_formulation}. We can also conclude that if the barrier conditions \eqref{eq:barrier_conditions_wdiff_params}, with parameters set according to Theorem \ref{thm:problem1_sol}, are always satisfied, then $\myvec{x}_j \in \myframe{V}_{\myframe{C}_t}, \ \forall t, \ \forall j$.

\begin{remark} \label{rmk:accessibility_of_sensors}
    The sensors (for a noiseless assumption) provide $\prescript{\myframe{C}_t}{}{}\myvec{x}_j$, thus we cannot strictly enforce \eqref{eq:barrier_conditions_wdiff_params} with parameters set according to Theorem \ref{thm:problem1_sol} since we have no access to $\prescript{\Tilde{\myframe{C}}_t}{}{}\myvec{x}_j$. 
\end{remark}

\begin{problem} \label{prob:fix}
Eqn. \eqref{eq:barrier_conditions_wdiff_params} can be written as $\myvec{\Theta}=\myvec{\Theta}(\prescript{\myframe{E}_t}{}{}\myvec{\myvec{u}}; \prescript{\hat{\myframe{C}}_t}{}{}\myvec{R}_{\myframe{C}_t}, \prescript{\hat{\myframe{C}}_t}{}{}\myvec{t}_{\myframe{C}_t}) \geq 0$. We seek to find $\Tilde{\myvec{\Theta}}(\prescript{\myframe{E}_t}{}{}\myvec{\myvec{u}}) \leq \myvec{\Theta}(\prescript{\myframe{E}_t}{}{}\myvec{\myvec{u}}; \prescript{\hat{\myframe{C}}_t}{}{}\myvec{R}_{\myframe{C}_t}, \prescript{\hat{\myframe{C}}_t}{}{}\myvec{t}_{\myframe{C}_t}) \ \forall \prescript{\hat{\myframe{C}}_t}{}{}\myvec{t}_{\myframe{C}_t}, \prescript{\hat{\myframe{C}}_t}{}{}\myvec{R}_{\myframe{C}_t}$ satisfying Assumption \ref{ass:general_problem}.
\end{problem}

$\myvec{\Theta}$ is decomposed into several terms: $\myvec{\Theta} =  \myvec{\Theta}_0(\prescript{\myframe{E}_t}{}{}\myvec{\myvec{u}}) + \myvec{\Theta}_1(\prescript{\myframe{E}_t}{}{}\myvec{\myvec{u}}, \prescript{\hat{\myframe{C}}_t}{}{}\myvec{t}_{\myframe{C}_t}) + \myvec{\Theta}_2(\prescript{\myframe{E}_t}{}{}\myvec{\myvec{u}}, \prescript{\hat{\myframe{C}}_t}{}{}\myvec{R}_{\myframe{C}_t}) +  \myvec{\Theta}_3( \prescript{\hat{\myframe{C}}_t}{}{}\myvec{R}_{\myframe{C}_t}) + \myvec{\Theta}_4(\prescript{\hat{\myframe{C}}_t}{}{}\myvec{t}_{\myframe{C}_t})$, where $\myvec{\Theta}_1(\prescript{\myframe{E}_t}{}{}\myvec{\myvec{u}}, \prescript{\hat{\myframe{C}}_t}{}{}\myvec{t}_{\myframe{C}_t}) = \myvec{\mu}_1^T(\prescript{\myframe{E}_t}{}{}\myvec{w}) \prescript{\hat{\myframe{C}}_t}{}{}\myvec{t}_{\myframe{C}_t}$ and $\myvec{\Theta}_2(\prescript{\myframe{E}_t}{}{}\myvec{\myvec{u}}, \prescript{\hat{\myframe{C}}_t}{}{}\myvec{R}_{\myframe{C}_t}) = \myvec{\mu}_1^T(\prescript{\myframe{E}_t}{}{}\myvec{w}) \prescript{\hat{\myframe{C}}_t}{}{}\myvec{R}_{\myframe{C}_t} \prescript{\myframe{C}_t}{}{}\myvec{x}_j$, with $\myvec{\mu}_1(\prescript{\myframe{E}_t}{}{}\myvec{w}) :=  -\prescript{\Tilde{\myframe{C}}_t}{}{}\myvec{R}_{\hat{\myframe{C}}_t}^T [\prescript{\myframe{E}_t}{}{}\myvec{R}_{\Tilde{\myframe{C}}_t} \prescript{\Tilde{\myframe{C}}_t}{}{}\Tilde{\myvec{a}}_i]_\times \prescript{\myframe{E}_t}{}{}\myvec{w}$, $\myvec{\Theta}_0(\prescript{\myframe{E}_t}{}{}\myvec{\myvec{u}}) = \boldsymbol{V}_i(\prescript{\myframe{E}_t}{}{}\myvec{\myvec{v}}) + \boldsymbol{W}_{ij}(\prescript{\myframe{E}_t}{}{}\myvec{\myvec{w}})$, $\myvec{\Theta}_3(\prescript{\hat{\myframe{C}}_t}{}{}\myvec{R}_{\myframe{C}_t}) = \alpha \prescript{\Tilde{\myframe{C}}_t}{}{}\Tilde{\myvec{a}}_i^T \prescript{\Tilde{\myframe{C}}_t}{}{}\myvec{R}_{\hat{\myframe{C}}_t} \prescript{\hat{\myframe{C}}_t}{}{}\myvec{R}_{\myframe{C}_t} \prescript{\myframe{C}_t}{}{}\myvec{x}_j$ and $\myvec{\Theta}_4(\prescript{\hat{\myframe{C}}_t}{}{}\myvec{t}_{\myframe{C}_t}) = \alpha \prescript{\Tilde{\myframe{C}}_t}{}{}\Tilde{\myvec{a}}_i^T \prescript{\Tilde{\myframe{C}}_t}{}{}\myvec{R}_{\hat{\myframe{C}}_t} \prescript{\hat{\myframe{C}}_t}{}{}\myvec{t}_{\myframe{C}_t}$. Problem \ref{prob:fix} can be solved by finding a lower bound to each term, $\Tilde{\myvec{\Theta}}_{k}(\prescript{\myframe{E}_t}{}{}\myvec{\myvec{u}}) \leq \myvec{\Theta}_k(\prescript{\myframe{E}_t}{}{}\myvec{\myvec{u}}; \prescript{\hat{\myframe{C}}_t}{}{}\myvec{R}_{\myframe{C}_t}, \prescript{\hat{\myframe{C}}_t}{}{}\myvec{t}_{\myframe{C}_t})$ for all $\prescript{\hat{\myframe{C}}_t}{}{}\myvec{t}_{\myframe{C}_t}$ and $\prescript{\hat{\myframe{C}}_t}{}{}\myvec{R}_{\myframe{C}_t}$ that satisfy Assumption \ref{ass:general_problem}. For the first term, it is straight forward that $\Tilde{\myvec{\Theta}}_0(\prescript{\myframe{E}_t}{}{}\myvec{\myvec{u}}) = \myvec{\Theta}_0(\prescript{\myframe{E}_t}{}{}\myvec{\myvec{u}})$ as $\myvec{\Theta}_0$ only depends on the control input. For the second term, it is clear that the minimizing $\prescript{\hat{\myframe{C}}_t}{}{}\myvec{t}_{\myframe{C}_t} = - \delta \myvec{\mu}_1(\prescript{\myframe{E}_t}{}{}\myvec{w}) / ||\myvec{\mu}_1(\prescript{\myframe{E}_t}{}{}\myvec{w})|| _2$ and $\Tilde{\myvec{\Theta}}_1(\prescript{\myframe{E}_t}{}{}\myvec{\myvec{u}}) = -\delta || \myvec{\mu}_1(\prescript{\myframe{E}_t}{}{}\myvec{w}) ||_2$, which is the biggest possible lower bound. For the third term, if we let $\myvec{r}_0 := \prescript{\myframe{C}_t}{}{}\myvec{x}_j \times \myvec{\mu}_1(\prescript{\myframe{E}_t}{}{}\myvec{w})$ with $\phi_0$ being the angle between $\prescript{\myframe{C}_t}{}{}\myvec{x}_j$ and $\myvec{\mu}_1(\prescript{\myframe{E}_t}{}{}\myvec{w})$, then the rotation vector minimizing $\prescript{\hat{\myframe{C}}_t}{}{}\myvec{R}_{\myframe{C}_t}$ is given by $-\epsilon \cdot \myvec{r}_0 / ||\myvec{r}_0||_2$ if $\phi_0 + \epsilon > \pi$ radians and $-||\prescript{\myframe{C}_t}{}{}\myvec{x}_j||_2 \cdot \myvec{\mu}_1 / ||\myvec{\mu}_1||_2$ else. $\Tilde{\myvec{\Theta}}_2(\prescript{\myframe{E}_t}{}{}\myvec{\myvec{u}})$ can then be computed using the Rodrigues formula where an arbitrary vector $v$ rotated by an angle $\myvec{\theta}$ around an axis with unit vector $k$ becomes
$v_{\text{rot}}=v \text{cos}(\myvec{\theta}) +(k \times v) \text{sin}(\myvec{\theta} + k (k \cdot v)(1-\text{cos}(\myvec{\theta}))$. For the forth term, if we let $\myvec{r}_1:= \prescript{\myframe{C}_t}{}{}\myvec{x}_j \times ( \prescript{\Tilde{\myframe{C}}_t}{}{}\myvec{R}_{\hat{\myframe{C}}_t}^T \prescript{\Tilde{C}_t}{}{}\Tilde{\myvec{{a}}}_i)^T / ||\prescript{\myframe{C}_t}{}{}\myvec{x}_j||_2 = \prescript{\myframe{C}_t}{}{}\myvec{x}_j \times \myvec{\mu}_2$ with $\phi_1$ being the angle between $\prescript{\myframe{C}_t}{}{}\myvec{x}_j$ and $\myvec{\mu}_2$, then the rotation vector minimizing $\prescript{\hat{\myframe{C}}_t}{}{}\myvec{R}_{\myframe{C}_t}$ is given by $-\epsilon \cdot \myvec{r}_1 / ||\myvec{r}_1||_2$ if $\phi_1 + \epsilon > \pi$ radians and $-||\prescript{\myframe{C}_t}{}{}\myvec{x}_j||_2 \cdot \myvec{\mu}_2 / ||\myvec{\mu}_2||_2$ else. $\Tilde{\myvec{\Theta}}_3(\prescript{\myframe{E}_t}{}{}\myvec{\myvec{u}})$ can be computed using again the Rodrigues formula. Finally, for $\myvec{\Theta}_4(\prescript{\hat{\myframe{C}}_t}{}{}\myvec{t}_{\myframe{C}_t})$, by the same process as before, the lower bound is $\Tilde{\myvec{\Theta}}_4 = - \alpha \delta$.

These constraints are minimally conservative, but non-convex in the control. To make the QP \eqref{eq:safety_set} solvable in real time with CBF constraints $\Tilde{\myvec{\Theta}}(\prescript{\myframe{E}_t}{}{}\myvec{\myvec{u}})$, we first solve the original QP with constraints \eqref{eq:vis_barrierconditions} and \eqref{eq:z_barrierconditions}, and then feed the result as an initial guess to a non-convex solver, which always succeeded experimentally. With this, we have created a robust visual servoing controller that ensures marker visibility.

\subsection{Human-In-the-Loop (HIL) control} \label{sec:hil}

The human can remotely control the end-effector velocity $\prescript{\myframe{E}_t}{}{}\boldsymbol{u}_{\text{HIL}}$, such that the nominal control is then given by 
\begin{equation} \label{eq:hil_nom}
\boldsymbol{u}_{\text{nom}} = \left(1 - \beta(h_{\text{min}})\right) \boldsymbol{u}_{\text{servo}} + \beta(h_{\text{min}}) \boldsymbol{u}_{\text{HIL}},    
\end{equation}
where $\beta(h_{\text{min}}) \in [0, 1]$ is the HIL relative importance. To compute it, we follow the adaptive rule in Algorithm \ref{alg:adaptive_rule}:

\begin{algorithm}
\caption{Adaptive rule for HIL relative importance $\beta$}\label{alg:adaptive_rule}
\begin{algorithmic}[1]
\State Compute $h_{\text{min}} = \min_{i,j} h_{ij}(\boldsymbol{x}_j)$.
\State  Define a safety threshold $h_{\text{safe}}$ from the visibility planes.
\State Adjust $\beta$: 
$\beta(h_{\text{min}}) = \beta_{\text{max}} \cdot sat\left( h_{\text{min}}/h_{\text{safe}} \right)$, where $\beta_{max} \in [0,1]$ and $sat(\cdot)$ is the saturation function
$sat(x) = \{
0 \ \text{if } x \leq 0, \
x \ \text{if } 0 < x < 1, \
1 \ \text{if } x \geq 1\}.$
\end{algorithmic}
\end{algorithm} 

\begin{corollary} 
Consider control law \eqref{eq:safety_set}, where $\myvec{u}_{\normalfont{\text{nom}}}$ is computed using \eqref{eq:hil_nom} with $\beta(h_{\normalfont{\text{min}}})$ given by Algorithm \ref{alg:adaptive_rule}. Then, the control input $\myvec{u}$ obtained by solving the QP \eqref{eq:safety_set} maximizes the allowable contribution of the human input $\myvec{u}_{\normalfont{\text{HIL}}}$ to the system control, while always ensuring marker visibility. 
\end{corollary}
\begin{proof} 
First, note that the QP \eqref{eq:safety_set} minimizes the difference between the actual control input $\myvec{u}$ and the nominal one $\myvec{u}_{\normalfont{\text{nom}}}$, while enforcing the CBF constraints for marker visibility. Following Definition \ref{def:cbf}, any $\myvec{u}$ satisfying \eqref{eq:cbf} renders the safe set forward invariant, guaranteeing marker visibility. In addition, $\myvec{u}_{\normalfont{\text{nom}}}$ in \eqref{eq:hil_nom} is a convex combination of the visual servoing control $\myvec{u}_{\normalfont{\text{servo}}}$ from \eqref{eq:visual_servoing_p_controller} and $\myvec{u}_{\normalfont{\text{HIL}}}$, weighted by $\beta(h_{\text{min}})$. Algorithm \ref{alg:adaptive_rule} adjusts $\beta$ based on the minimum barrier functions, i.e., when $h_{\text{min}} \leq 0$, the marker is at or beyond the visibility boundary, so we set $\beta=0$ to rely entirely on $\myvec{u}_{\normalfont{\text{servo}}}$ and regain visibility. When $0 < h_{\text{min}} < h_{\text{safe}}$, the marker is approaching the boundary, so we smoothly decrease $\beta$ to reduce the influence of $\myvec{u}_{\normalfont{\text{HIL}}}$ and increase it for $\myvec{u}_{\normalfont{\text{servo}}}$. Lastly, when $h_{\text{min}} \geq h_{\text{safe}}$, the marker is at a safe distance from the visibility planes so we set $\beta=\beta_{max}$, allowing maximum influence of $\myvec{u}_{\normalfont{\text{HIL}}}$. By design, this rule ensures $\myvec{u}_{\normalfont{\text{nom}}}$ has as much of $\boldsymbol{u}_{\text{HIL}}$ as safely possible, given the current state of the system and the constraints. Since the QP \eqref{eq:safety_set} seeks to find $\myvec{u}$ as close as possible to $\myvec{u}_{\text{nom}}$ satisfying the CBF constraints, it effectively maximizes the allowable influence of $\boldsymbol{u}_{\text{HIL}}$ while always ensuring marker visibility. 
\end{proof}

\section{Experimental Results}

\subsection{Software architecture}

% The software developed is a set of ROS2 \cite{ros2} nodes, shown in Fig. \ref{fig:software_architecture}, that provide elementary types of actions, with more complex planning built on top. The code leverages ROS2 libraries and two specific nodes of the MoveIT2 library \cite{sucan_chitta_moveit}. The \textsc{Move\_group} node stores information about the physical model of the robots and their joint limits, tracks their current state using encoders and the Qualysis motion capture system \cite{qualysis}, and keeps a planning scene with obstacles. It computes inverse kinematics (IK) through damped least squares \cite{damped_least_squares}, performs obstacle free motion planning by employing BFMT* \cite{BFMT_star} and computes forward kinematics. Only one robotic platform runs \textsc{Move\_group} while the other ones communicate via WiFi with it. \textsc{Servo\_node} receives twist commands from \eqref{eq:safety_set} for the end-effector, computes the corresponding joint displacements through the damped least squares algorithm for IK, and publishes them to the hardware for execution. It also decelerates when the arm is close to a collision or a singularity. Each robotic platform runs its own instance of \textsc{Servo\_node}. The \textsc{ros2\_aruco} package is used for ArUco detection \cite{aruco_paper} and pose estimation with the Intel Realsense D435i camera \cite{realsense_camera}. Finally, a handheld controller is used to send HIL commands.

\begin{figure}[ht]
\centering
\includegraphics[width=\linewidth]{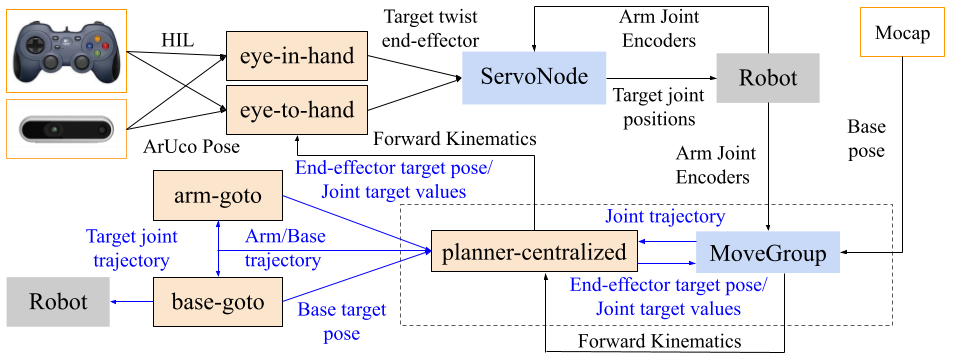}
\caption{Information flow diagram of the developed actions.} 
\label{fig:software_architecture}
\end{figure}

The software, shown in Fig. \ref{fig:software_architecture}, consists of ROS2 nodes providing elementary actions for higher-level planning. The code leverages ROS2 libraries and two specific nodes of the MoveIT2 library \cite{sucan_chitta_moveit}. The \textsc{Move\_group} node stores information about the physical model of the robots and their joint limits, tracks their current state using encoders and the Qualysis motion capture system, and keeps a planning scene with obstacles. It computes inverse kinematics (IK) through damped least squares \cite{damped_least_squares}, performs obstacle free motion planning by employing BFMT* \cite{BFMT_star} and computes forward kinematics. Only one robot runs \textsc{Move\_group}; the others communicate via WiFi, each running \textsc{Servo\_node} locally. \textsc{Servo\_node} receives twist commands from \eqref{eq:safety_set} for the end-effector, computes joint displacements through the damped least squares algorithm for IK, and publishes them to the hardware. It also decelerates the arm close to collisions or singularities. The \textsc{ros2\_aruco} package is used for pose estimation \cite{aruco_paper} with the Intel Realsense D435i camera. Finally, a joystick is used to send HIL commands.

\subsection{Results}
\begin{figure}[ht]
    \centering
    % First column with two images stacked, labeled (a)
    \begin{minipage}{0.5\linewidth}
        \centering
        \includegraphics[width=0.893\linewidth]{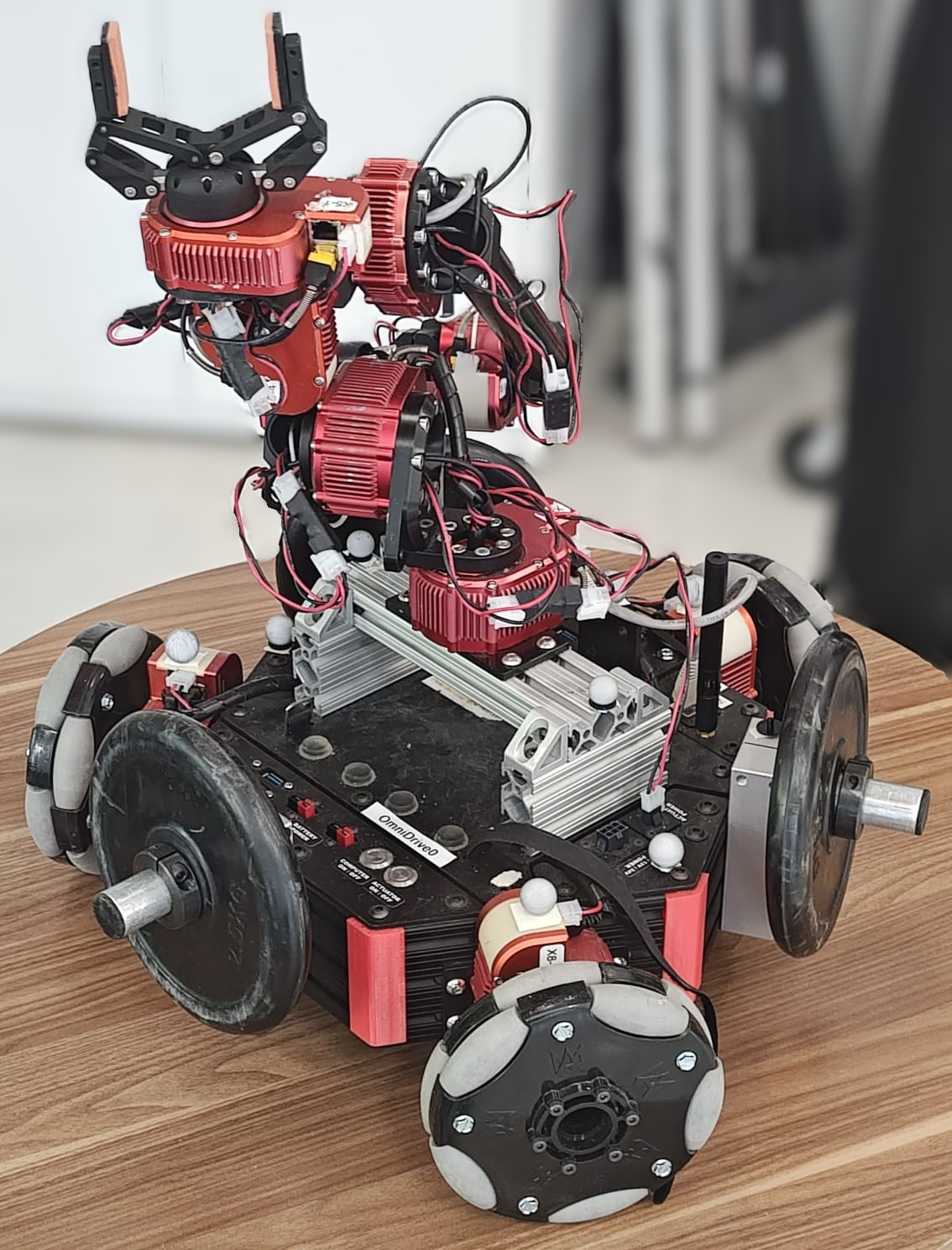}
    \end{minipage}%
    % Second column with one image, labeled (b)
    \begin{minipage}{0.5\linewidth}
        \centering
        \includegraphics[width=0.875\textwidth]{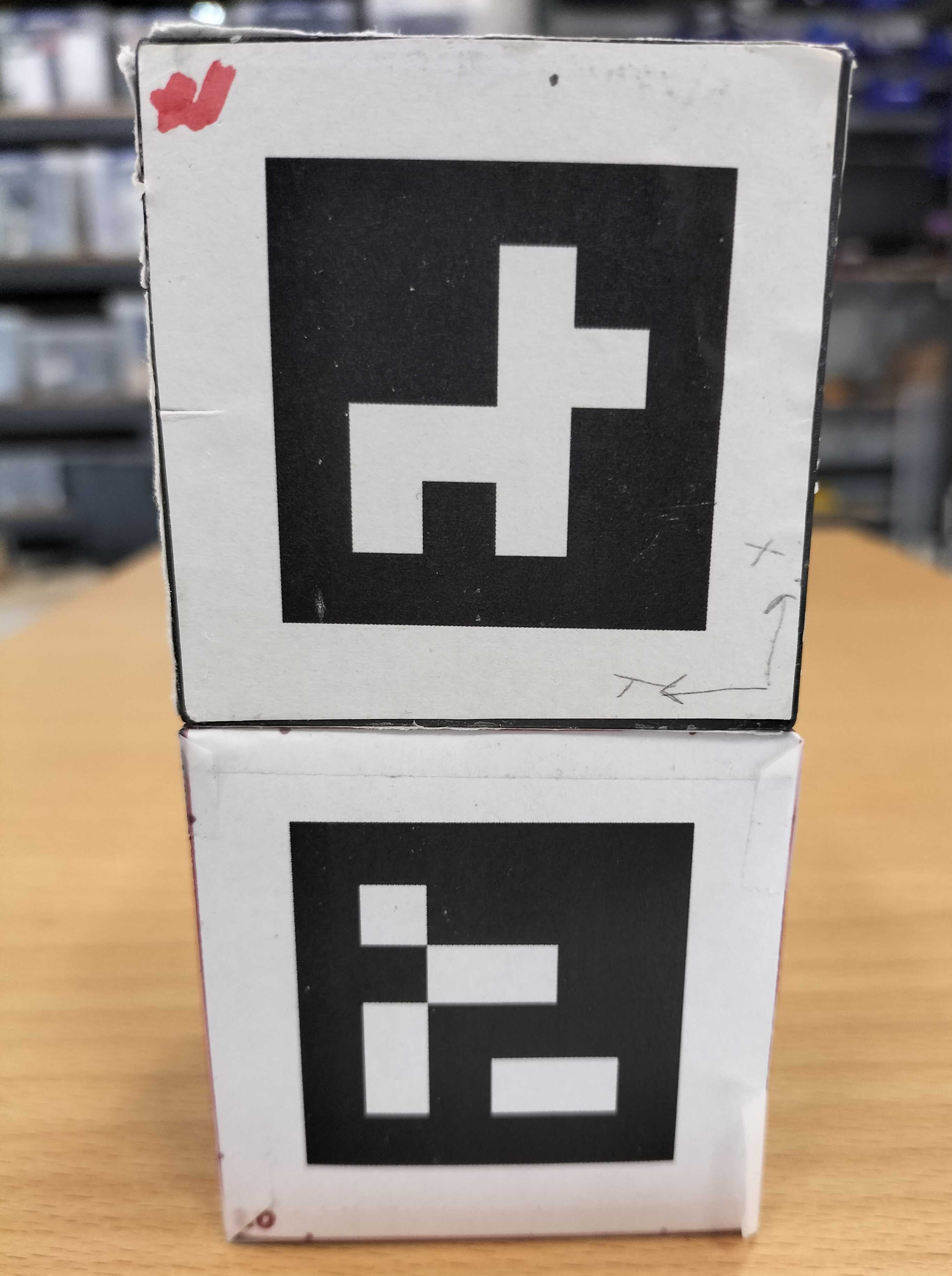}
    \end{minipage}
    \caption{Experimental setup for assembly tasks using the (left) Hebi Rosie mobile manipulators and the (right) cubes with ArUco markers.}
    \label{fig:exp_workspace}
\end{figure}

Here we show the results of two experiments related to construction assembly. The experimental setup, shown in Fig. \ref{fig:exp_workspace}, consists of a few objects marked with ArUcos resting on top of static surfaces, assuming 2cm/5° calibration uncertainty, and one or two robots starting in front of it with at least one of the markers in the field of view of the camera. The video of both experiments can be found in \cite{video}.

% \begin{figure}[ht]
% \centering
% \includegraphics[width=\linewidth]{figures/exp_workspace.jpg}
% \caption{Experimental setup}
% \label{fig:exp_workspace}
% \end{figure}

\subsection{Servoing with HIL} \label{subsec:servoing_wHIL}
\begin{figure}[ht]
\centering
\includegraphics[width=0.9\linewidth]{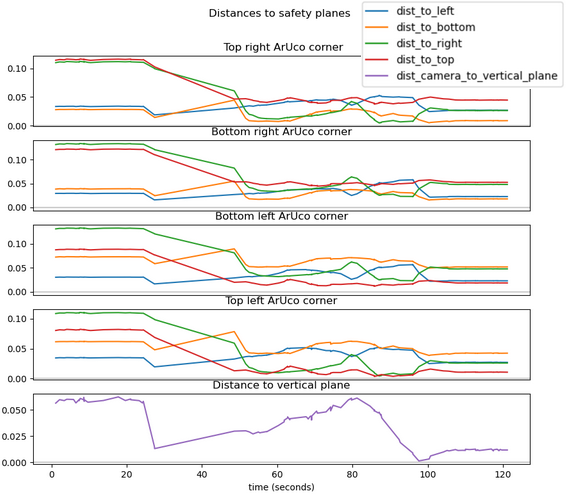}
\caption{Time evolution of the CBFs with a grey line indicating zero value.} 
\label{fig:cbfs_wHIL}
\end{figure}

We show the efficacy of the framework in keeping marker visibility. The CBF constraints (i.e., the distances of each ArUco corner with the visibility planes), shown in Fig. \ref{fig:cbfs_wHIL}, always remain non-negative, showing that visibility was never lost. In \cite{video}, we also show an example of a translation and rotation of the end-effector with and without the CBFs. As expected, when the CBFs are deactivated, some corners of the ArUco go out of the field of view of the camera. Lastly, \cite{video} demonstrates two HIL examples showing safety-constrained human control and a simple pick-and-place task.

\subsection{Building a three block tower}
In this experiment, two Hebi robots collaborate to build a structure consisting of three stacked cubes. Their starting position is another tower that the robots need to disassemble first. One robot acts as the mover, picking and placing the cubes and, since handling the objects leads to camera occlusions, the other robot provides visual feedback about the relative pose of the handled object and the structure. As shown in \cite{video}, the structure is completed successfully, always ensuring marker visibility, precision and structural integrity.

\section{Conclusion}

In this work, we implemented a set of actions that allow to precisely pick and place objects. We proposed CBFs to ensure fiducial markers in the object remain detectable throught the robot movement and robustified them to account for calibration errors. The algorithms were also extended to allow the system to be operable by a human. We showed consistent, successful assembly tasks, where CBFs effectively maintained object visibility. Future work will focus on markerless object pose estimation, CBFs for dynamic objects and more advanced multi-robot collaborative manipulation.

%%%%%%%%%%%%%%%%%%%%%%%%%%%%%%%%%%%%%%%%%%%%%%%%%%%%%%%%%%%%%%%%%%%%%%%%%%%%%%%%

% \newpage
 
\bibliographystyle{IEEEtran}
\bibliography{IEEEabrv,references}

\end{document}